\documentclass[12pt]{article}

\usepackage{amsmath}
\usepackage{graphicx}
\usepackage{amssymb}
\usepackage{amsfonts}
\usepackage{latexsym}
\def\beq{\begin{eqnarray}}
\def\eeq{\end{eqnarray}}
\def\ln{\,\mbox{ln}\,}

\renewcommand{\vec}[1]{{\bf #1}}

\begin{document}
\begin{titlepage}
\title{Scalar Field Cosmologies With Inverted Potentials}
\author{B. Boisseau$^1$\thanks{email:bruno.boisseau@lmpt.univ-tours.fr}, 
H. Giacomini$^1$\thanks{email:hector.giacomini@lmpt.univ-tours.fr},
D. Polarski$^2$\thanks{email:david.polarski@umontpellier.fr},\\
\hfill\\
$^1$Universit\'e de Tours, Laboratoire de Math\'ematiques et Physique Th\'eorique,\\
CNRS/UMR 7350, 37200 Tours, France\\
$^2$Universit\'e Montpellier \& CNRS, Laboratoire Charles Coulomb,\\
UMR 5221, F-34095 Montpellier, France}

\pagestyle{plain}
\date{\today}

\maketitle

\begin{abstract}
Regular bouncing solutions in the framework of a scalar-tensor gravity model were 
found in a recent work. We reconsider the problem in the Einstein frame (EF) 
in the present work. Singularities arising at the limit of physical viability 
of the model in the Jordan frame (JF) are either of the Big Bang or of the Big 
Crunch type in the EF. As a result we obtain integrable scalar field cosmological 
models in general relativity (GR) with inverted double-well potentials unbounded 
from below which possess solutions regular in the future, tending to a de Sitter space, 
and starting with a Big Bang. The existence of the two fixed points for the field 
dynamics at late times found earlier in the JF becomes transparent in the EF. 

\end{abstract}
\end{titlepage}

%%%%%%%%%%%%%%%%%%%%%%%%%%%%%%%%%%%%%%%%%%%%%%%%%%%%%%%%%%%%%%%%%%%%%%%%%%%%%%%%%%%%%%%%%%%%%%
%%%%%%%%%%%%%%%%%%%%%%%%%%%%%%%%%%%%%%%%%%%%%%%%%%%%%%%%%%%%%%%%%%%%%%%%%%%%%%%%%%%%%%%%%%%%%%
%%%%%%%%%%%%%%%%%%%%%%%%%%%%%%%%%%%%%%%%%%%%%%%%%%%%%%%%%%%%%%%%%%%%%%%%%%%%%%%%%%%%%%%%%%%%%%

\section{Introduction}
The possibility to produce bouncing universes has attracted a lot of interest over the years.  
As Friedman-Lema\^itre-Robertson-Walker (FLRW) universe possess generically an initial 
singularity, the possibility to avoid it in this way has become even more challenging.
The simplest example where a bouncing universe is obtained and which has a nonzero measure in 
the space of initial conditions is that of a massive scalar field in a closed FLRW universe
~\cite{Star78} where the curvature singularity is generically moved to the past. Non-singular 
solutions can be constructed but they are degenerate (i.e. for a set of initial conditions 
which is of measure zero) \cite{Page84}, see also \cite{Ka98}.
However a bounce with a positive spatial curvature requires severe fine tuning of 
initial conditions before the contraction stage \cite{Star78}, \cite{GT08}.
Spatially-flat FLRW non-degenerate bouncing universes have been built outside general 
relativity like theories with scalar~\cite{DSNA11,BS13} or tensor ghosts, loop quantum 
gravity (see e.g. \cite{APS06}) or gravity described by an effectively non-local 
Lagrangian~(see e.g.\cite{QECZ11},\cite{ESV11}, and \cite{BP14} for a recent review).

Recently viable non-degenerate bouncing solutions were found in the framework of ghost-free 
scalar-tensor gravity in a spatially-flat FLRW universe \cite{BGPS15}. The construction of a 
bouncing universe in General Relativity (GR) requires violation of the weak-energy condition 
and this is well-known to be allowed in scalar-tensor gravity \cite{BEPS00},\cite{GPRS06}. 
Indeed, the Friedmann equations can be written as in Einstein gravity but now with an 
effective component of the phantom type. Nevertheless, it came as a surprise that such a 
well-behaved and intensively investigated extension of General Relativity (GR) like scalar-
tensor gravity allowed for this family of non-degenerate spatially-flat bouncing universes. 
The model considered amounts to a conformally coupled scalar field and a quartic self 
interaction potential in Einstein gravity with a positive cosmological constant \cite{BGPS15}. 
It is interesting that the conformal invariance was arrived at due to some mathematical 
requirements imposed on the coupled equations of motion. While this specific model has been 
investigated in different contexts assuming metrics different from our FRLW metric 
\cite{MTZ03}, \cite{HPP06}, the possibility to produce bouncing solutions was not 
considered before. 

It is well-known that any scalar-tensor model in the Jordan frame can be expressed as a 
mathematically equivalent problem in the Einstein frame (EF) where gravity is described by 
General Relativity, and some new potential and a non-minimal coupling of matter to 
gravity arise. This is why consideration of this theory in the EF often turns out to be 
enlightening. While it was clear from a direct inspection of the problem in the JF that 
two cases had to be considered either with an inverted potential (when the field kinetic 
term in the lagrangian is positive, $Z=1$) or with a potential bounded from below when $Z=-1$, 
we will see that both cases are very similar when viewed in the EF with the appearance in 
both cases of an inverted double-well potential $V$.  

Two critical points for the dynamics of the scalar field $\Phi$ at late times were found 
while inspection of the Jordan frame (JF) potential did not offer any clue as to their 
existence. So, while the viable solutions were found, their physical significance remained 
unclear to some extent. As we will show, the existence of these two critical points become 
transparent in the EF frame. 
Further, interesting behaviours arise for the scale factor in this frame. 
Our EF analysis will yield integrable scalar field spatially-flat FLRW universes whose 
singularities are of the Big Bang or of the Big Crunch type. 
The paper is constructed as follows: In Section 2, we present the bouncing model in the JF. 
In Section 3, we study the problem in the EF with a 
detailed study of solutions corresponding to all possible bouncing universes in the JF. 
Finally our findings and conclusions are summarized in Section 4. 

%%%%%%%%%%%%%%%%%%%%%%%%%%%%%%%%%%%%%%%%%%%%%%%%%%%%%%%%%%%%%%%%%%%%%%
\section{A bouncing model}
%%%%%%%%%%%%%%%%%%%%%%%%%%%%%%%%%%%%%%%%%%%%%%%%%%%%%%%%%%%%%%%%%%%%%%

We consider a universe with gravity described by a scalar-tensor theory. The 
Lagrangian density in the Jordan frame of the gravitational sector is given by 
\beq
\label{LJF}
L = \frac{1}{2} \left[ F(\Phi)R - Z(\Phi)~g^{\mu\nu}\partial{\mu}\Phi\partial{\nu}\Phi 
                                              - 2 U(\Phi) \right]~.
\eeq
We will use below the freedom to take $Z=1$ or $Z=-1$, corresponding 
physically to $\omega_{BD}>0$ or $\omega_{BD}<0$ ($\omega_{BD} = 
Z F \left(\frac{dF}{d\Phi}\right)^{-2}$). 
When $\omega_{BD}<0$, the theory is ghost-free provided $-\frac{3}{2}<\omega_{BD}<0$.  
For spatially flat FLRW universes with metric $ds^2=-dt^2+a^2(t)d\vec{x}^2$, the modified 
Friedmann equations read 
\beq
-3 F H^2 + \frac12 Z~\dot{\Phi}^2 - 3 H \dot{F} + U &=& 0~, \label{Fr1}\\
2 F \dot{H} + Z~\dot{\Phi}^2 + \ddot{F} - H \dot{F} &=& 0~, \label{Fr2}
\eeq
with $H\equiv \frac{\dot{a}}{a}$. Here and below a dot, resp. a prime, stands for the 
derivative with respect to $t$, resp. to $\Phi$. 
The equation of motion of $\Phi$ 
\beq
Z~(-\ddot{\Phi} - 3 H \dot{\Phi}) + \frac{R}{2}F' - U' = 0~, \label{ddPhi}
\eeq
is contained in \eqref{Fr1},\eqref{Fr2}.
We take the following ansatz
\beq
Z F &=& -\frac{1}{6} \Phi^2 + \kappa^{-2}, \label{F}\\
Z U &=& \frac{\Lambda}{\kappa^2} - c \Phi^4 ~,\label{U}
\eeq
where $\kappa^{-2}>0$, $\Lambda>0$ and $c$ are constant parameters, only 
$c$ being dimensionless. 
Equations \eqref{Fr1}-\eqref{ddPhi} are invariant under the transformation $Z=1\to Z=-1$, 
however the domain of validity $F>0$ is changed, so that we have two different problems. 
It is crucial to realize that with \eqref{F} $\Phi$ becomes a conformally coupled scalar field 
with an additional quartic self-interaction $-c\Phi^4$.   
Interestingly, we first arrived at this model by requiring that 
a combination of \eqref{Fr1}, \eqref{Fr2} and \eqref{ddPhi}, without derivation with 
respect to $t$, be zero. Due to its underlying symmetry, this model has been considered in 
the past in completely different contexts (see e.g. \cite{MTZ03}, \cite{HPP06}).
%Therefore, the energy-momentum tensor of the field $\Phi$ is the sum of ``dark'' 
%radiation with a negative energy density and a cosmological constant.

The system \eqref{Fr1}, \eqref{Fr2} reduces to
\beq
3H^2 = \Lambda + \kappa^2 \frac{A}{a^4}, \label{Eq1}\\
\frac{1}{2}\left(\frac{d\chi}{d\eta}\right)^2 - c\chi^4 = A~, \label{Eq2}
\eeq
with an effective positive cosmological constant $\Lambda$ and the bare gravitational 
constant $8\pi G \equiv \kappa^2$, while $\chi = a \Phi$ and 
$\eta=\int dt/a(t)$. 
The crucial point is that the constant $A$ -- the energy density of the 
field $\chi$ in Minkowski space-time -- can be negative. An effective phenomenological 
model in the framework of GR with \eqref{Eq1} was considered in \cite{P13}

Scalar-tensor models can accommodate an effective dark energy component of the 
phantom type ($w_{eff}<-1$) \cite{BEPS00},\cite{GPRS06} and this is precisely 
what we have for $A<0$. In this case the second term on the right hand side of 
\eqref{Eq1} can be seen as corresponding to dark radiation. 
Remarkably, our system can be completely integrated and the analytical expression 
for $\Phi(t)$ when $\dot{\Phi}_0 < 0$ 
\beq
\Phi(t) = \frac{ -\left( \frac{\Lambda}{c \kappa^2}\right)^{\frac{1}{4}} }
{\sqrt{\cosh(2 \sqrt{\frac{\Lambda}{3}}~t)}}  \frac{1}{{\rm dn}\left({\rm dn}^{-1}
\left(-\left( \frac{\Lambda}{c \kappa^2}\right)^{\frac{1}{4}} \frac{1}{\Phi_0}~|~2 \right)  
-i\sqrt{2}(\frac{9c}{\kappa^2\Lambda})^{\frac{1}{4}}
                F\left(i\sqrt{\frac{\Lambda}{3}}~t~|~2\right)~|~2\right) } \label{Phit}
\eeq
is given in terms of the Jacobi elliptic function ${\rm dn}(u|2)$ (${\rm dn}^{-1}(u|2)$ 
stands for its inverse) and of the 
the elliptic integral of the first kind $F(x|2)$ \cite{AbrSte}, and $\Phi_0$ is the 
field value at the bounce. 
 
Bouncing solutions are obtained for $A<0$ and from 
\eqref{Eq2} we have $c>0$ so that $U$ is necessarily an \emph{inverted} potential, 
unbounded from below when $Z=1$. Though this may look unphysical at first sight, scalar fields 
with such an interaction have been often considered both in quantum field theory and 
cosmology, see e.g.~\cite{Rub09}. Our present analysis in the EF reinforces our 
belief that such potentials should not be ruled out a priori.   
Integrating \eqref{Eq1}, a bouncing solution is obtained 
\beq
a = a_0 \cosh^{\frac{1}{2}}\Big[2 \sqrt{ \frac{\Lambda}{3} }~t\Big]~, \label{a}
%H &=& \sqrt{ \frac{\Lambda}{3} } \tanh \Big[2 \sqrt{\frac{\Lambda}{3}}~t\Big]~,\label{H}
\eeq
where 
\beq
a_0=\left(\frac{-A \kappa^2}{\Lambda}\right)^{\frac{1}{4}}
\eeq
is the value of $a$ at the bounce located at $t=0$ with a trivial redefinition of $t$. 
It satisfies $\dot{H} > 0$ and has a constant Ricci scalar $R=6(\dot{H}+2H^2)= 4 \Lambda$.
So we see that $a(t)$ is integrated independently of $\Phi(t)$. The coupling of the two 
equations of motion is through the integration constant $A$.
Solving for the wave equation rewritten as follows
\beq
\ddot{\Phi} + 3 H \dot{\Phi} + 4 c \Phi ( \tilde{\Phi}^2 - \Phi^2 ) = 0~,\label{ddPhi2}
\eeq
where $\tilde{\Phi}\equiv (\frac{\Lambda}{6c})^{\frac{1}{2}}$, subject to the constraint 
\eqref{Eq2} at the bounce 
\beq
\frac12 \dot{\Phi}_0^2  - c ( \Phi_0^4 - \Phi_{0,min}^4 ) = 0~, \label{ic}
\eeq
where $\Phi_{0,min}\equiv (\frac{\Lambda}{\kappa^2 c})^{\frac{1}{4}} = 
(\frac{-A}{c a^4_0})^{\frac{1}{4}}$, our system is solved. 

The requirement $F>0$ implies $\Phi < \frac{\sqrt{6}}{\kappa}$ when $Z=1$, and 
$\Phi > \frac{\sqrt{6}}{\kappa}$ when $Z=-1$. 
Note that $\Phi=0$ at a finite time is impossible from \eqref{Eq2} also for $Z=1$.
From \eqref{Fr1}, the condition $U(\Phi_0)\equiv U_0\le 0$ for $Z=1$ 
gives immediately $\Phi_0\ge \Phi_{0,min}$, while $\Phi < \frac{\sqrt{6}}{\kappa} 
\equiv\Phi_{max}$ is valid at all times.
A non vanishing interval for $\Phi_0$ requires $\Phi_{0,min} < \Phi_{max}$ or 
\beq
\frac{\Lambda~\kappa^2}{36c} < 1~, ~~~~~~~~~~~~~~~~~~~~~~~~~Z=1~, \label{parcond}
\eeq
implying in turn $\tilde{\Phi}<\Phi_{0,min}$.
Hence, choosing $\Phi_0 > 0$, the following ordering is obtained for $Z=1$
\beq
\tilde{\Phi} < \Phi_{0,min} \leq \Phi_0 <  \Phi_{max}~. \label{int2}
\eeq
When $Z=-1$, the condition $U_0\ge 0$ coming from \eqref{Fr1} gives again 
$\Phi_0\ge \Phi_{0,min}$ while $\Phi > \frac{\sqrt{6}}{\kappa}$ holds at all times. 
Imposing the inequality 
\beq
\frac{\Lambda~\kappa^2}{36c} > 1~, ~~~~~~~~~~~~~~~~~~~~~~~~~~~~~~Z=-1~, \label{parcondb}
\eeq
the following ordering is obtained 
\beq
\frac{\sqrt{6}}{\kappa}  < \Phi_{0,min} < \tilde{\Phi} ~, \label{int2b}
\eeq
and 
\beq
\Phi_{0,min} \leq \Phi_0~. \label{int2bb}
\eeq
Finally, one can show that for bouncing universes $\dot{\Phi}=0$ is possible in the 
intervals 
\beq
\tilde{\Phi} \le \Phi \le \Phi_{0,min}~,~~~~~~~~~~~~~~~~~~~~~~~~~~~~&&\dot{\Phi}=0,
                                                      ~\ddot{\Phi}>0,~Z=1~, \label{in1}\\
\Phi_{0,min} \le \Phi \le \tilde{\Phi}~,~~~~~~~~~~~~~~~~~~~~~~~~~~~~&&\dot{\Phi}=0,
                                                      ~\ddot{\Phi}<0,~Z=-1~.     \label{in-1}  
\eeq
Indeed from \eqref{Fr1} when $\dot\Phi = 0$, we easily get 
\beq
-c~( \Phi^4 - \Phi_{0,min}^4 ) + \frac{H^2}{2}~( \Phi^2 - \Phi^2_{max}) = 0~.\label{eq1}
\eeq 
As for $Z=1$ we have $\Phi<\Phi_{max}$ hence \eqref{eq1} requires $\Phi \le \Phi_{0,min}$ 
(the inequality saturates when $H=0$). 
For $Z=-1$ however, we have $\Phi>\Phi_{max}= \frac{\sqrt{6}}{\kappa}$ so that \eqref{eq1} 
requires $\Phi \ge \Phi_{0,min}$ in that case. 
On the other hand, using \eqref{Fr2} and \eqref{ddPhi}, we obtain that $\dot\Phi = 0$ 
is possible only for $\Phi\ge \tilde\Phi$ when $Z=1$ and for $\Phi\le \tilde\Phi$ when 
$Z=-1$ so that \eqref{in1}, \eqref{in-1} are immediately obtained. 
Remember that for $Z=-1$ we have $\tilde\Phi > \Phi_{0,min}$.
The inequalities \eqref{in1}, \eqref{in-1}, will be very useful in the analysis of the 
field dynamics in the EF. 

We recall the essential findings in \cite{BGPS15} concerning the JF dynamics. 
For $Z=1$, solutions with 
$\dot{\Phi}_0 < 0,~\Phi_{0,cr} < \Phi_0 < \Phi_{max}$ tend asymptotically to zero 
while solutions satisfying $\dot{\Phi}_0 < 0,~\Phi_{0,min} < \Phi_0 < \Phi_{0,cr}$ 
reach $\Phi_{max}$ (and actually diverge) in a finite time and are therefore 
unviable as they leave the interval for which $F>0$. There is a critical value 
$\Phi_{0,cr}$ separating the two behaviours \cite{BFG07} and solutions starting from 
$\Phi_{0,cr}$ with $\dot{\Phi}_0 < 0$ tend to $\tilde{\Phi}$. All solutions with 
$\dot{\Phi}_0 > 0$ will diverge in the future in a finite time. As a corollary, even 
regular solutions for $t>0$ will necessarily diverge at a finite negative time as 
the equations are invariant under time inversion. As the equations are symmetric 
under the transformation $\Phi\to -\Phi$, it is enough to consider $\Phi>0$. 

For $Z=-1$, all solutions with $\Phi_0 > \tilde{\Phi}$ are unviable, they tend either 
to zero for $\dot{\Phi}_0 < 0$ (therefore crossing $\frac{\sqrt{6}}{\kappa}$ where 
$F$ vanish before changing sign), or they tend to $\infty$ in a finite time for 
$\dot{\Phi}_0 > 0$. The only viable solution has 
$\dot{\Phi}_0 > 0,~\Phi_0 = \Phi_{0,cr} < \tilde{\Phi}$ and it tends to $\tilde{\Phi}$. 
Solutions with $\Phi_0 < \Phi_{0,cr}$ will tend to zero, whatever the sign of $\dot{\Phi}_0$. 
Solutions with $\Phi_0 > \Phi_{0,cr}$ tend to $\infty$ for $\dot{\Phi}_0 > 0$, and to 
zero for $\dot{\Phi}_0 < 0$. This more complicated behaviour for $Z=-1$ will be easily 
visualized in our EF analysis.

%%%%%%%%%%%%%%%%%%%%%%%%%%%%%%%%%%%%%%%%%%%%%%%%%%%%%%%%%%%%%%%%%%%%%%%%%%%
%%%%%%%%%%%%%%%%%%%%%%%%%%%%%%%%%%%%%%%%%%%%%%%%%%%%%%%%%%%%%%%%%%%%%%%%%%%
\section{The problem in the Einstein frame (EF)}
%%%%%%%%%%%%%%%%%%%%%%%%%%%%%%%%%%%%%%%%%%%%%%%%%%%%%%%%%%%%%%%%%%%%%%%%%%%
%%%%%%%%%%%%%%%%%%%%%%%%%%%%%%%%%%%%%%%%%%%%%%%%%%%%%%%%%%%%%%%%%%%%%%%%%%%
It is well-know that model \eqref{LJF} can be expressed in the Einstein frame (EF) where 
the lagrangian becomes
\beq
L = \frac{R_*}{2\kappa^2} - \frac{1}{2}g_*^{\mu\nu}~\partial_{\mu}\phi \partial_{\nu}\phi 
                                   - V(\phi)~.\label{LEF}
\eeq 
An asterisk denotes expressions in the so-called EF with metric $g_{*,\mu\nu}$. One can go 
from one frame to the other with the following transformations (see e.g. \cite{EP00})
\footnote{Our conventions differ from those in \cite{EP00} as here scalar fields in both 
frames are not dimensionless.} 
\beq
\left( \frac{d\phi}{d\Phi} \right)^2 &=& \frac{2}{\kappa^2} ~\left[ \frac{3}{4} 
     \left( \frac{dF/d\Phi}{F} \right)^2 + \frac{Z}{2F} \right]~, \label{dphi} \\
g_{*,\mu\nu} &=& \kappa^2 ~F(\Phi) ~g_{\mu\nu}~,\\
V(\phi) &=& \kappa^{-4} ~U(\Phi) ~F^{-2}(\Phi)~.\label{V}
\eeq
Matter in the EF is gravitationally coupled to the JF metric $g_{\mu\nu} = A^2(\phi) ~g_{*,\mu\nu}$ 
with 
\beq
A^2(\phi) = \kappa^{-2} ~F^{-1}(\Phi)~.\label{A}   
\eeq
We have a flat FLRW universe in the EF as well, viz.
\beq
ds_*^2=-dt_*^2 + a_*^2(t_*) ~d\vec{x}^2~,
\eeq
where $t_*$, resp. $a_*$, is the EF time, resp. the scale factor, defined as 
\beq
dt &=& A(\phi)~dt_*~,  \label{tt*} \\
a &=& A(\phi)~a_*~.  \label{aa*}
\eeq
Using \eqref{dphi} and \eqref{tt*} we can relate the field derivatives in both frames 
and we obtain 
\beq
\dot{\Phi} =  \frac{d\Phi}{d\phi}~A^{-1}(\phi)~\frac{d\phi}{dt_*}~.\label{dphidPhi}
\eeq 
Let us turn now to the EF dynamics. The corresponding Friedmann equations become
\beq
3~H_*^2 &=& \kappa^2 \left[\frac{1}{2} \left( \frac{d\phi}{dt_*}\right)^2 + V \right]~,\label{F1}\\
\frac{dH_*}{dt_*} &=& - \frac{\kappa^2}{2} \left( \frac{d\phi}{dt_*} \right)^2~,\label{F2}
\eeq
and finally the EF scalar field $\phi$ obeys the usual Klein-Gordon equation in the absence 
of matter
\beq
\frac{d^2\phi}{dt_*^2} + 3~H_*~\frac{d\phi}{dt_*} + \frac{dV}{d\phi} = 0~. \label{KG}
\eeq
Equation \eqref{KG} can be obtained from \eqref{F1}, \eqref{F2}. We note that with our 
conventions, equations governing the dynamics in the EF reduce to those of a minimally 
coupled scalar field with potential $V$ in general relativity.   

It is easy to relate the Hubble parameter in both frames and we find
\beq
H = A^{-1}(\phi)~\left( H_* + \frac{d\ln A(\phi)}{d\phi}~\frac{d\phi}{dt_*} \right)~. 
\label{HH*} 
\eeq
Hence, we see from \eqref{HH*} that the existence of a bounce in the JF implies  
\beq
H_{*,0} = -\frac{d\ln A(\phi)}{d\phi} \frac{d\phi}{dt_*}\Big|_0~,\label{H*0}
\eeq
where we have chosen $t_*(t=0)=0$ (see below), i.e. the bouncing time $t=0$ in the JF 
corresponds to the EF time $t_*=0$.  
In particular for $H_*$ and $\frac{d\ln A}{d\phi}$ positive at $t_*=0$, we must have 
$\frac{d\phi}{dt_*}\big|_0 < 0$. 
Expression \eqref{HH*} can be used in order to relate the dynamics in both frames.

As for any scalar field cosmology, we can start at some arbitrary initial time $t_i$ with 
the three arbitrary initial values, $a_{*,i}$, $\phi_i$ and 
$\left(\frac{d\phi}{dt_*}\right)_i$. Then, from \eqref{F1} we get $\dot{a}_{*,i}$ and we 
can just evolve the system. 
In this way however nothing guarantees that the solution in the EF correspond to a 
bouncing solution in the JF. 
Hence, it will be convenient for our discussion to look at the problem in a slightly different 
way. We will choose the EF time $t_*$ in such a way that $t_*=0$ corresponds to the bouncing 
time $t=0$ in the JF. This corresponds to the choice of one integration 
constant. We can choose freely two more initial values, and we choose $a_{*,0}$, $\phi_0$. 
The field value $\phi_0$ must satisfy a condition corresponding to $\Phi_0\ge \Phi_{0,min}$ 
and this will be easily implemented in the EF as we will see below.  
Making use of \eqref{aa*}, we find $a_0$ and therefore also the integration constant 
$A\equiv - \frac{a_0^4~\Lambda}{\kappa^2}$ appearing in \eqref{Eq1}, \eqref{Eq2}. 
Note that an arbitrary choice of $a_{*,0}$ will correspond 
to a choice of the (negative) integration constant $A$. At $t_*=0$ \eqref{H*0} holds, hence 
$H_{*,0}$ is a function of $\phi_0$ and $\frac{d\phi}{dt_*}\big|_0$. Writing \eqref{Eq2} in 
the EF variables, $\frac{d\phi}{dt_*}\big|_0$ is given (up to a sign) and also $H_{*,0}$. 
To summarize, with our choice of time $t_*(t=0)=0$, starting at $t_*=0$ and choosing 
$a_{*,0}$ and $\phi_0$, we can construct all the solutions corresponding to bouncing solutions 
in the JF.  
%%%%%%%%%%%%%%%%%%%%%%%%%%%%
%%%%%%%%%%%%%%%%%%%%%%%%%%%%
\subsection{The case Z=1}
%%%%%%%%%%%%%%%%%%%%%%%%%%%%
%%%%%%%%%%%%%%%%%%%%%%%%%%%%
For our model \eqref{F}, \eqref{U}, using \eqref{dphi}, \eqref{V}, \eqref{A}, we find
\beq
\phi &=& \frac{\sqrt{6}}{\kappa}~{\rm arctanh} \left(\frac{\kappa}{\sqrt{6}}\Phi 
                                                             \right)~,\label{phi}\\ 
\Phi &=& \frac{\sqrt{6}}{\kappa}~\tanh \left(\frac{\kappa}{\sqrt{6}}\phi \right)~.\label{Phi}
\eeq
The transformation $\Phi(\phi)$ is monotonically growing hence the ordering in both 
frames between two values is unchanged. We have from the general transformation equations 
\beq
A(\phi) &=& \cosh \left(\frac{\kappa}{\sqrt{6}}\phi \right)~,\\
V(\phi) &=& \frac{\Lambda}{\kappa^2} 
    \left[ \cosh^4 \left( \frac{\kappa}{\sqrt{6}} \phi \right) - 
 \frac{36 c}{\kappa^2 \Lambda}~\sinh^4 \left( \frac{\kappa}{\sqrt{6}} \phi \right) \right]~.
\label{V1}
\eeq
Surprisingly, the potential \eqref{V1} corresponds to an \emph{inverted} double-well potential. 
We note further that $V(\phi)\to \frac{\Lambda}{\kappa^2}$ for $\phi\to 0$.
%%%%%%%%%%%%%%%%%%%%%%%%%%%%%%%%%%%%%%%%%%%%%%%%%%%%%%%%%%%%%
%%%%%%%%%%%%%%%%%%%%%%%%%%%%%%%%%%%%%%%%%%%%%%%%%%%%%%%%%%%%%
\begin{figure}
\begin{centering}
\includegraphics[scale=.7]{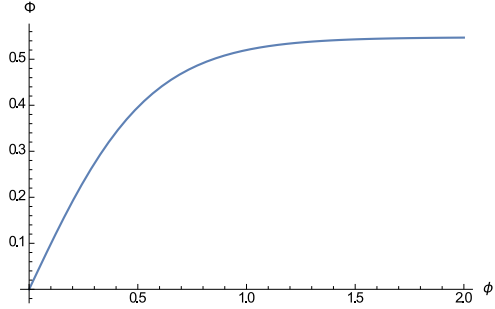}
\par\end{centering}
\caption{The JF field $\Phi$ is shown as a function of the EF field $\phi$ for $Z=1$ 
with the parameters $\kappa=\sqrt{20},~c=3$ and $\Lambda=5$. 
It is seen that $\Phi$ is a monotonically growing function of $\phi$. Therefore 
the ordering $0 < \tilde\Phi < \Phi_{0,min} < \frac{\sqrt{6}}{\kappa}\equiv \Phi_{max}$ is 
preserved in EF, namely $0 < \tilde\phi < \phi_{0,min} < \infty$. The limit 
$\Phi\to \frac{\sqrt{6}}{\kappa}\equiv \Phi_{max}$ corresponds to $\phi\to \infty$. 
The interval $0 < \phi < \infty$ covers the physically viable interval 
$0 < \Phi < \Phi_{max}$ for which $F>0$ in the JF .} 
\label{figPhi1}
\end{figure}
%%%%%%%%%%%%%%%%%%%%%%%%%%%%%%%%%%%%%%%%%%%%%%%%%%%%%%%%%%%%%
%%%%%%%%%%%%%%%%%%%%%%%%%%%%%%%%%%%%%%%%%%%%%%%%%%%%%%%%%%%%%

%%%%%%%%%%%%%%%%%%%%%%%%%%%%%%%%%%%%%%%%%%%%%%%%%%%%%%%%%%%%%
%%%%%%%%%%%%%%%%%%%%%%%%%%%%%%%%%%%%%%%%%%%%%%%%%%%%%%%%%%%%%
\begin{figure}
\begin{centering}
\includegraphics[scale=.7]{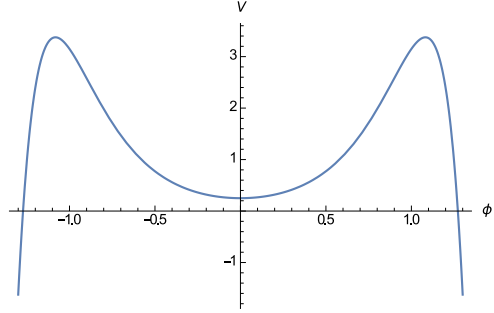}
\par\end{centering}
\caption{The EF potential $V$ is shown in the case $Z=1$ for the parameters of Figure 
\ref{figPhi1}. The potential $V$ has a maximum 
in $\tilde\phi = \frac{\sqrt{6}}{\kappa}~{\rm arctanh} 
\big[ \left( \frac{\kappa^2 \Lambda}{36 c} \right)^{\frac{1}{2}} \big]$ and it vanishes at 
$\phi_{0,min} = \frac{\sqrt{6}}{\kappa}~{\rm arctanh} 
\big[ \left(\frac{\kappa^2 \Lambda}{36 c} \right)^{\frac{1}{4}} \big] $. 
%they correspond respectively to the points $\tilde\Phi$ and $\Phi_{0,min}$ in the JF. 
The value $\phi=0$ corresponds to $\Phi= 0$ in the JF and $\phi=\infty$ corresponds 
to $\Phi = \frac{\sqrt{6}}{\kappa}\equiv \Phi_{max}$, the unphysical limit where $F$ vanishes 
and for which either a Big Bang or a Big Crunch takes place in the EF.  
In contrast to the JF, $\tilde\phi$ gets a direct meaning looking at $V$. 
Bouncing solutions in the JF satisfy $\phi_0 \ge\phi_{0,min}$ in the EF, with the bouncing 
time $t=0$ corresponding to $t_*=0$.}
\label{figV1}
\end{figure}
%%%%%%%%%%%%%%%%%%%%%%%%%%%%%%%%%%%%%%%%%%%%%%%%%%%%%%%%%%%%%
%%%%%%%%%%%%%%%%%%%%%%%%%%%%%%%%%%%%%%%%%%%%%%%%%%%%%%%%%%%%%

The physical regime in the JF corresponds to $\Phi < \Phi_{max} = \frac{\sqrt{6}}{\kappa}$. 
Using \eqref{phi} to find the coresponding value in the EF we get  
\beq
\phi_{max} = \frac{\sqrt{6}}{\kappa}~{\rm arctanh} \left( 1 \right) = \infty~.\label{phimax}
\eeq
Hence $\phi$ can be arbitrarily large. 
From \eqref{tt*}, the EF time $t_*$ and the JF time $t$ are related as follows
\beq
t_*(t) = \int_0^t \sqrt{ 1 - \left(\frac{\Phi(t')}{\Phi_{max}}\right)^2 }~dt'~,\label{t*(t)}
%t_*(t) = \int_0^t \frac{dt'}{\cosh \left( \frac{\kappa}{\sqrt{6}} \phi(t') \right)}~,
\eeq 
where the integration constant is choosen such that the EF time $t_*$ and the Jordan time 
$t$ vanish simultaneously. It is crucial to note that $t_*$ is a monotonically growing function 
of $t$ hence both times are essentially equivalent. We have also 
\beq
a = \cosh \left(\frac{\kappa}{\sqrt{6}}\phi \right)~a_*~.\label{aa*1}
\eeq
We easily derive the following equality
\beq
\frac{d\phi}{dt_*} = \cosh^3\left( \frac{\kappa}{\sqrt{6}}\phi \right)~\dot{\Phi}~,
\label{dPhi}
\eeq 
which shows that the field derivatives have the same sign in both frames and also that 
the asymptotic behaviour $\frac{d\Phi}{dt}\to 0,~\Phi\to 0$ for $t\to \infty$ in the JF 
corresponds to $\frac{d\phi}{dt_*}\to 0,~\phi\to 0$ for $t_*\to \infty$ in the EF.
We further find from \eqref{HH*} that
\beq
H =  \frac{1}{\cosh \left(\frac{\kappa}{\sqrt{6}}\phi \right)}~\left( H_* + 
   \frac{\kappa}{\sqrt{6}}~\tanh \left(\frac{\kappa}{\sqrt{6}}\phi \right)~
                               \frac{d\phi}{dt_*} \right) ~.\label{HH*1} 
\eeq
For viable bouncing solutions in the JF $\Phi$ decreases monotonically and tends either to 
$\tilde{\Phi}\equiv (\frac{\Lambda}{6c})^{\frac{1}{2}}$ or to zero for $t\to \infty$.  
From \eqref{t*(t)}, $t\to \infty$ corresponds to $t_*\to \infty$. The value $\tilde{\Phi}$ 
corresponds to $\tilde\phi$ in the EF with
\beq
\tilde{\phi} = \frac{\sqrt{6}}{\kappa}~{\rm arctanh}
        \left[ \left( \frac{\kappa^2 \Lambda}{36 c} \right)^{\frac{1}{2}} \right]~.\label{tilphi}
\eeq
It is easily checked that $V$ has a maximum at $\phi=\tilde{\phi}$. 
While the value $\tilde\Phi$ in the JF cannot be understood from inspection 
of the potential $U(\Phi)$, in the EF on the contrary it has an obvious meaning. 
The potential $U$ vanishes at $\Phi_{0,min}\equiv (\frac{\Lambda}{\kappa^2 c})^{\frac{1}{4}}$, 
and the potential $V$ will vanish at the corresponding value $\phi_{0,min}$ given by
\beq  
\phi_{0,min} = \frac{\sqrt{6}}{\kappa}~{\rm arctanh} 
     \left[ \left(\frac{\kappa^2 \Lambda}{36 c} \right)^{\frac{1}{4}} \right]~.\label{phimin}
\eeq
We start with general considerations concerning the appearance of singularities in the EF. 
From \eqref{F2} $H_*$ decreases monotonically in sharp contrast with the behaviour 
$\dot{H}>0$ in the JF. At $\Phi\to \Phi_{max}$ ($F\to 0$), the model would become unphysical. 
We have seen in \cite{BGPS15} that $\Phi_{max}$ is reached in a finite time $t$ while 
$\dot{\Phi}$ does not tend to zero there and so from \eqref{dPhi}, $\frac{d\phi}{dt_*}$ goes 
to infinity because $\phi\to \infty$. Moreover $a_*\to 0$ in a finite time $t_*$ from 
\eqref{aa*} as $A(\phi)\to \infty$. Therefore we see from \eqref{F1} that $H_*$ diverges 
because from \eqref{dPhi} the kinetic term dominates the potential term for $\phi\to \infty$. 
To summarize, when $\Phi=\Phi_{max}$ in the JF, we get either a Big Bang or a Big Crunch 
type singularity in the EF, depending on the sign of $H_*$. 
This will also happen  at some finite time $t<0$ for bouncing solutions regular in the future. 
Actually it is well-known that expansion in one frame can correspond to a contraction in 
the other frame (see e.g. \cite{GV03}) and this is what takes place here. 

We will consider now all the bouncing solutions, either viable or unviable, found in our 
earlier analysis in the JF in \cite{BGPS15} and study their behaviour in the EF. 
We note first that these solutions for $Z=1$ must satisfy $\phi_0 \ge \phi_{0,min}$. Further, 
from \eqref{in1}, \eqref{phi}, $\frac{d\phi}{dt_*}=0,~\frac{d^2\phi}{dt_*^2}>0$ is possible 
only in the interval $\tilde{\phi} \le \phi \le \phi_{0,min}$. 

We consider first the solution with $\phi_0 = \phi_{0,min}$. This solution has 
$\Phi_0 = \Phi_{0,min},~\dot\Phi_0=0$ in the JF. We have from \eqref{dPhi} that 
$\frac{d\phi}{dt_*}\big|_0=0$ in the EF too, and hence also $H_*=0$ from \eqref{HH*1}. 
This is also clear from \eqref{F1} because $\phi_{0,min}$ is the value where $V$ vanishes. 
From \eqref{F2} $H_*<0$ for $t_*>0$. As in the JF we have $\dot{\Phi}(t>0)>0$ for this 
solution, this must also be the case in the EF, i.e. $\frac{d\phi}{dt_*}(t_*>0)>0$. 
So the field $\phi$ tends to $\infty$ while the universe contracts reaching $a_*=0$ in a 
finite time. For $t<0$, this solution started in the JF from 
$\Phi_{max}$ at some finite negative time with $\dot{\Phi}(t<0)<0$. This corresponds in the 
EF to a solution starting with a Big Bang at $\phi=\infty$, reaching a maximal expansion at 
$\phi_{0,min}$ with $\phi$ reaching its minimum, and eventually recontracting and ending in a 
Big Crunch with $\phi\to \infty$. It is the only bouncing solution in the JF for which the 
Hubble parameters and the field derivatives vanish simultaneously in both frames. 

Let us consider next the solutions with $\Phi_{0,min} < \Phi_0 < \Phi_{0,cr}$ and 
$\dot{\Phi}_0 < 0$. We know that in the JF, these solutions will diverge, crossing 
the value $\Phi_{max}$ in a finite time. After the bounce, these solutions have first 
$\dot{\Phi} < 0$ before reaching a minimum. As we have shown earlier, this will happen 
in the interval $\tilde{\Phi}<\Phi<\Phi_{0,min}$ in the JF.
As the fields $\Phi$ and $\phi$ reach a turning point simultaneously in both frames, 
in the EF this corresponds to $\phi_0$ somewhere in the negative part of the potential $V$, 
entering the positive part of $V$ and turning back at some point 
$\tilde{\phi} < \phi < \phi_{0,min}$ before increasing and tending to infinity in a finite 
time $t_*$. 
At the bounce we have from \eqref{HH*1}, $H_*>0$ because $H=0$ while 
$\dot{\Phi}_0<0$ or $\frac{d\phi}{dt_*}\big|_0<0$. 
At the minimum of $\Phi(t)$ and $\phi(t_*)$, we have from \eqref{HH*1} that $H_*$ has the 
same sign as $H$, which is positive for $t>0$.
As long as we are in the positive part of $V$, $H_*=0$ is impossible from \eqref{F1}. 
On the other hand we see from \eqref{aa*} that $a_*\to 0$ as $\phi\to \infty$ 
because the corresponding JF time is finite. 
This means that in the EF, the universe must recontract. So $H_*$ must first vanish, 
which is possible for $\phi > \phi_{0,min}$, before changing sign. 
Again, it was found that these solutions would inevitably 
reach the value $\Phi=\Phi_{max}$ at some finite negative time in the past with 
$\dot{\Phi}(t<0)<0$.
This corresponds again in the EF to a solution starting from $\phi=\infty$ with a Big Bang 
and $\frac{d\phi}{dt_*}<0$. 
To summarize, in all these cases a universe starting with a Big Bang and ending in a Big 
Crunch is obtained. 

We consider now the bouncing solution in the JF tending to $\tilde{\Phi}$ and starting 
precisely from $\Phi_{0,cr}$. This corresponds to a solution starting from $\phi_{0,cr}\equiv 
\frac{\sqrt{6}}{\kappa}~{\rm arctanh} \left(\frac{\kappa}{\sqrt{6}}\Phi_{0,cr} \right)$ 
with $\frac{d\phi}{dt_*}\big|_0<0$ in the EF. This solution has exactly the kinetic energy 
required in order to reach $\tilde{\phi}$ at the top of the potential $V(\tilde\phi)$, 
for $t_*\to \infty$. 
This is obviously an unstable fixed point of the system, found earlier from our analysis in 
the JF, which can be reached only starting from $\phi_{0,cr}$. Note that the asymptotic 
Hubble parameters differ in both frames in this case.  

Finally, the solutions with $\phi_0 > \phi_{0,cr}$ 
will have lower initial negative velocity $\frac{d\phi}{dt_*}\big|_0$ (larger initial 
kinetic energy) with enough kinetic energy to climb up the potential and pass its top 
after which $\phi$ cannot stop and $\phi=0$ will be reached after an infinite time $t_*$. 
This corresponds to an overdamped regime where the absence of 
oscillations is due to the (strong) friction term  in \eqref{KG}. 
This can be shown in either frame, let us do this in the JF. 

At $t\to \infty$ and $\Phi\to 0$, \eqref{ddPhi2} gives 
\beq
\ddot{\Phi} + 2\lambda~\dot{\Phi} + \omega^2~\Phi = 0~,\label{dosc}
\eeq
with $\lambda\equiv \frac{\sqrt{3\Lambda}}{2},~\omega^2\equiv \frac{2\Lambda}{3}$. Clearly, as 
we have $\lambda > \omega$ we obtain an overdamped regime without oscillations where $\Phi$ 
vanishes exponentially. The asymptotic behaviour to leading order of $\Phi$ is 
\beq
\Phi \sim c_1~\exp \left( -\sqrt{\frac{\Lambda}{3}}~t \right)~.\label{Phias}
\eeq
We have also $\phi\to 0$ and $\frac{d\phi}{dt_*}\to 0$ for $t_*\to \infty$.
Using \eqref{Phi}, \eqref{t*(t)}, $\phi(t_*)$ has no oscillations either. 
From \eqref{F1}, \eqref{V1}, the universe tends asymptotically to the same de Sitter space 
in both frames with $3H_*^2 = 3H^2 = \Lambda$.
If $\frac{d\phi}{dt_*}\big|_0 > 0$, $\phi\to \infty$ in a finite time $t_*$ and 
a Big Crunch is obtained   

We want to conclude this subsection with another important aspect of our results. 
Substituting \eqref{Phit} in \eqref{phi}, using \eqref{aa*1} and inverting 
\eqref{t*(t)} in order to find $t(t_*)$, the Friedmann equation in GR for a spatially flat 
cosmology and for a minimally coupled scalar field with potential \eqref{V1} is completely 
integrated. Though $V$ is an inverted potential, even unbounded from below, 
regular solutions in the future do exist as we have shown, starting with a Big Bang singularity 
in the past. Other non viable solutions (in the JF) exhibit a Big Crunch which interestingly 
appears as well in integrable scalar field cosmologies for non inverted potentials bounded 
from below with negative extrema \cite{FSS13}. The integrable potentials mentioned there are 
combinations of exponentials, which is also the case for our model. 
% and it is interesting that these non negative potentials seem to arise naturally in $N=1$ 
% or extended SUGRA \cite{FSS13}. 

%%%%%%%%%%%%%%%%%%%%%%%%%%%%
%%%%%%%%%%%%%%%%%%%%%%%%%%%%
\subsection{The case Z=-1}
%%%%%%%%%%%%%%%%%%%%%%%%%%%%
%%%%%%%%%%%%%%%%%%%%%%%%%%%%
Let us consider now the case $Z=-1$ where both functions in \eqref{F} and \eqref{U} are 
multiplied by $-1$.  This case shows subtle differences with the case $Z=1$. 
The following transformation is obtained 
\beq
\phi &=& \frac{\sqrt{6}}{2\kappa} ~\ln \left[ \frac{\Phi + \frac{\sqrt{6}}{\kappa} }
  {\Phi - \frac{\sqrt{6}}{\kappa}} \right]~,\label{phiPhib}\\
\Phi &=& \frac{\sqrt{6}}{\kappa}~\coth \left(\frac{\kappa}{\sqrt{6}} \phi \right)~,
\label{Phiphib}
\eeq
where we have taken into account $\Phi > \frac{\sqrt{6}}{\kappa}$. 
In sharp contrast to the case $Z=1$, $\Phi$ is now a monotonically \emph{decreasing} 
function of $\phi$, hence the ordering of corresponding values in both frames is inverted 
when we go from one frame to the other. 
%%%%%%%%%%%%%%%%%%%%%%%%%%%%%%%%%%%%%%%%%%%%%%%%%%%%%%%%%%%%%
%%%%%%%%%%%%%%%%%%%%%%%%%%%%%%%%%%%%%%%%%%%%%%%%%%%%%%%%%%%%%
\begin{figure}
\begin{centering}
\includegraphics[scale=.7]{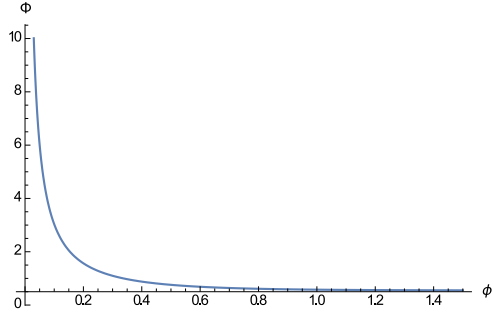}
\par\end{centering}
\caption{The JF field $\Phi$ is shown as a function of the EF field $\phi$ for $Z=-1$ with 
the parameters $\kappa=\sqrt{20},~c=3$ and $\Lambda=\frac{13}{2}$. 
It is seen that $\Phi$ is now a monotonically decreasing function of $\phi$ with 
$\Phi\to \infty$ when $\phi\to 0$ and $\Phi\to \frac{\sqrt{6}}{\kappa}$ ($0.5477$ with these 
parameters) for $\phi\to \infty$. 
Hence the ordering $\frac{\sqrt{6}}{\kappa} < \Phi_{0,min} < \tilde\Phi < \infty$ is 
inverted and becomes in the EF, $\infty > \phi_{0,min} > \tilde\phi > 0$. Note that this 
ordering in the EF is the same for both cases $Z=1$ and $Z=-1$. The interval 
$0 < \phi < \infty$ in the EF corresponds to the physically viable interval 
$\frac{\sqrt{6}}{\kappa} < \Phi < \infty$ in the JF for which $F>0$.}
\label{figPhi-1}
\end{figure}
%%%%%%%%%%%%%%%%%%%%%%%%%%%%%%%%%%%%%%%%%%%%%%%%%%%%%%%%%%%%%
%%%%%%%%%%%%%%%%%%%%%%%%%%%%%%%%%%%%%%%%%%%%%%%%%%%%%%%%%%%%%

Remember that the allowed range for $\Phi$ in the JF satisfying $F>0$ corresponds to 
$\Phi > \frac{\sqrt{6}}{\kappa}$. When $\Phi\to \frac{\sqrt{6}}{\kappa}$, we have 
$\phi\to \infty$ while $\Phi\to \infty$ corresponds to $\phi\to 0$. 
We have further from \eqref{A}
\beq
A(\phi) &=& \sinh \left(\frac{\kappa}{\sqrt{6}} \phi \right)~, \label{Aphi-1}\\
t_*(t) &=& \int_0^t \sqrt{ \left(\frac{\Phi(t')}{\sqrt{6}/\kappa}\right)^2 - 1 }~dt'~,\label{t*(t)b}\\
%t_*(t) &=& \int_0^t \frac{dt'}{\sinh \left( \frac{\kappa}{\sqrt{6}} \phi(t') \right)}~,
a &=& \sinh \left(\frac{\kappa}{\sqrt{6}}\phi \right)~a_*~.\label{aa*b}
\eeq
For the potential $V$, one obtains
\beq
V(\phi) = \frac{36 c}{\kappa^4}~\left[ \cosh^4 \left(\frac{\kappa}{\sqrt{6}} \phi \right) - 
     \frac{\kappa^2 \Lambda}{36 c} \sinh^4 \left(\frac{\kappa}{\sqrt{6}} \phi \right) \right]~.
\label{Vb}
\eeq
%%%%%%%%%%%%%%%%%%%%%%%%%%%%%%%%%%%%%%%%%%%%%%%%%%%%%%%%%%%%%
%%%%%%%%%%%%%%%%%%%%%%%%%%%%%%%%%%%%%%%%%%%%%%%%%%%%%%%%%%%%%
\begin{figure}
\begin{centering}
\includegraphics[scale=.7]{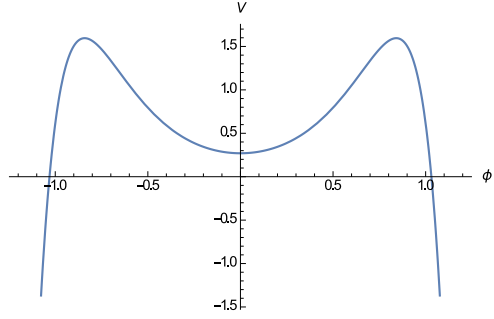}
\par\end{centering}
\caption{The EF potential $V$ is shown in the case $Z=-1$ and the same parameters as in 
Figure \ref{figPhi-1}. 
The potential $V$ has a maximum in $\tilde\phi = \frac{\sqrt{6}}{\kappa}~{\rm arccoth} 
\big[ \left( \frac{\kappa^2 \Lambda}{36 c} \right)^{\frac{1}{2}} \big]$ and it vanishes at 
$\phi_{0,min} = \frac{\sqrt{6}}{\kappa}~
{\rm arccoth} \big[ \left(\frac{\kappa^2 \Lambda}{36 c} \right)^{\frac{1}{4}} \big]$, 
corresponding to the points $\tilde\Phi$, respectively  
$\Phi_{0,min}$, in the JF. The point $\phi=0$ corresponds now to $\Phi=\infty$.
Again, the unphysical limit $F=0$ is pushed at $\phi_{max}=\infty$, 
%corresponding to $\Phi = \frac{\sqrt{6}}{\kappa}\equiv \Phi_{max}$ in the JF, 
where again either a Big Bang or a Big Crunch takes place in the EF. 
Note the striking analogy with the case $Z=1$. 
Bouncing solutions in the JF satisfy $\phi_0 \le\phi_{0,min}$ in the EF.}
\label{figV-1}
\end{figure}
%%%%%%%%%%%%%%%%%%%%%%%%%%%%%%%%%%%%%%%%%%%%%%%%%%%%%%%%%%%%%
%%%%%%%%%%%%%%%%%%%%%%%%%%%%%%%%%%%%%%%%%%%%%%%%%%%%%%%%%%%%%

As for $Z=-1$ the inequality $\frac{\kappa^2 \Lambda}{36 c} > 1$ is assumed in order to 
ensure a nonvanishing interval of field values $\Phi_0$ at the bounce (see \eqref{IntJF-1}), 
the second (negative) term inside the brackets in \eqref{Vb} will dominate when 
$\phi\to \infty$ yielding as in the case $Z=1$ an inverted double-well potential 
%tending to $-\infty$ for $\phi\to \infty$. 
in sharp contrast to the JF potential $U$ which is not inverted in this case. 
It is easily checked that we have here too a 
maximum at $\phi=\tilde\phi$ defined as
\beq
\coth \left[\frac{\kappa}{\sqrt{6}} \tilde\phi \right] = 
                  \left( \frac{\kappa^2 \Lambda}{36c} \right)^{\frac{1}{2}}~,
\eeq
while $V$ is zero at $\phi=\phi_{0,min}$ given by 
\beq
\coth \left[\frac{\kappa}{\sqrt{6}} \phi_{0,min} \right] = 
                  \left( \frac{\kappa^2 \Lambda}{36c} \right)^{\frac{1}{4}}~.
\eeq
As we have in the JF 
\beq
\frac{\sqrt{6}}{\kappa} < \Phi_{0,min} < \tilde{\Phi}~,\label{IntJF-1}
\eeq
we get in the EF  
\beq
 \tilde{\phi} < \phi_{0,min} < \infty~,\label{IntEF-1}
\eeq
hence the ordering \eqref{IntEF-1} in the EF is similar to the EF ordering for $Z=1$, 
while the ordering in the JF for $Z=1$ and $Z=-1$ is inverted. 
This arises because the transformation $\eqref{phiPhib}$ or $\eqref{Phiphib}$ invert again 
this ordering so that finally the ordering in the EF is the same while it gets inverted in 
the JF. 
It is further easily checked from \eqref{dphidPhi}, \eqref{Phiphib}, that for $Z=-1$ the field 
derivatives in both frames have opposite signs, namely
\beq
\frac{d\phi}{dt_*} = -\sinh^3 \left(\frac{\kappa}{\sqrt{6}} \phi \right) ~\dot{\Phi}~.
\eeq
From the point of view of the potential $V$, a situation very similar to the case $Z=1$ is 
recovered. Despite the fact that the potential $U$ in the JF is not inverted for $Z=-1$, the 
corresponding potential $V$ in the EF is inverted and analogous to the one obtained for $Z=1$. 
This is a consequence of the negative kinetic term in the JF when $Z=-1$.  

We now turn our attention to the study of all bouncing solutions in the JF for $Z=-1$ found in 
\cite{BGPS15}. In all cases we must have $\phi_0 < \phi_{0,min}$ (this corresponds to $U>0$ in 
the JF). From \eqref{in-1}, \eqref{phiPhib}, $\frac{d\phi}{dt_*}=0,~\frac{d^2\phi}{dt_*^2}>0$ 
is posible only in the interval $\tilde{\phi} \le \phi \le \phi_{0,min}$. 
The EF picture is similar to $Z=1$ in this respect. 

We can start in the positive region of $V$ either on the left of $\tilde{\phi}$ or on the 
right. 
\par\noindent
a) We start with $\phi_0 < \tilde{\phi}$ and $\frac{d\phi}{dt_*}\big|_0 < 0$ 
corresponding to $\Phi_0 > \tilde{\Phi}$ and $\dot{\Phi}_0 > 0$.
It was found in the JF that $\Phi\to \infty$ in a finite time $t$, which amounts to 
$\phi\to 0$ in an infinite time $t_*$. 
The fact that $\phi\to 0$ is obvious looking at the initial conditions and at the shape 
of $V$. Note that $a_*$ diverges too there. This case was rejected in the JF because 
$\Phi\to \infty$ in a finite time $t$. The corresponding 
EF time is pushed to $t_*=\infty$. The EF dynamics cannot go beyond the physically valid 
regime in the JF. Hence if we postulate that the physical time is the EF time $t_*$ then this 
solution is perfectly valid. However if we postulate that the physical frame is the JF, then 
the physical time is the JF time $t$ and this solution must be discarded.  

When $\phi_0 < \tilde{\phi}$ and $\frac{d\phi}{dt_*}\big|_0 > 0$ corresponding to 
$\Phi_0 > \tilde{\Phi}$ and $\dot{\Phi}_0 < 0$, it was found that $\Phi\to 0$ and hence 
$\Phi\to \frac{\sqrt{6}}{\kappa}$ in a finite time $t$. This corresponds $\phi\to \infty$, 
again for a finite time $t_*$.  
This dynamics is again easily understood in the EF.
As $\phi$ cannot stop and turn back 
%neither at values lower than $\tilde{\phi}$ nor at values larger than $\tilde{\phi}$ 
because this would correspond to a maximum (which is impossible, see \eqref{in-1}), it will 
pass $\tilde{\phi}$ and roll down towards 
infinity. Note that if some solution starting from $\phi_{0,cr}$ was allowed to stop precisely 
at $\tilde{\phi}$, it would mean that solutions around it would still go to infinity, 
which is impossible because $\phi_{0,cr}$ must separate different behaviours of the late 
time dynamics. So this is impossible and it shows also that $\phi_{0,cr}>\tilde{\phi}$. 
Finally as we have $a_*\to 0$, this case will end in a Big Crunch. 

\par\noindent
b) We consider now $\phi_0 > \tilde{\phi}$ and $\frac{d\phi}{dt_*}\big|_0 > 0$, or
$\Phi_0 < \tilde{\Phi}$ and $\dot{\Phi}_0 < 0$. In that case, it was shown that 
$\Phi\to \frac{\sqrt{6}}{\kappa}$ in a finite time $t$. 
Hence $\phi\to \infty$ in a finite time $t_*$ and here too a Big Crunch is obtained in the EF. 
This dynamics is very easily understood by inspection of $V$: 
the system starts somewhere in the interval  $\tilde{\phi} < \phi_0 < \phi_{0,min}$ and moves 
downwards along the potential. 

We consider again $\phi_0 > \tilde{\phi}$ but now $\frac{d\phi}{dt_*}\big|_0 < 0$.
($\Phi_0 < \tilde{\Phi}$ and $\dot{\Phi}_0 > 0$). Three possible behavious were found in the 
JF which are easily represented in the EF: 
\par\noindent
1) $\phi\to \infty$, $a_*\to 0$ in a finite time $t_*$ so a Big Crunch is obtained. This 
corresponds to $\Phi\to 0$ and  $F\to 0$ in a finite time. So $\phi$ climbs up the 
potential, stops before $\tilde\phi$ and then rolls down to infinity. This solution was 
rejected in the JF because the condition $F>0$ would eventually be violated after a 
finite time and in the EF this corresponds to a Big Crunch. 
\par\noindent
2) $\phi\to 0$ after an infinite time $t_*$ or $\Phi\to \infty$ diverges in a finite time 
$t$. In this case, $\phi$ has enough kinetic energy to pass $\tilde\phi$ and as it not 
allowed to stop afterwards, it eventually settles down in $\phi=0$ after an infinite time 
$t_*$. This is reminiscent to a case considered earlier: the EF dynamics is perfectly 
acceptable if we agree that the EF time $t_*$ is the physical time and not the JF time $t$. 
\par\noindent
3) Finally, we have the only viable solution found in the JF for which 
$\Phi\to \tilde{\Phi}$ in the asymptotic future. This corresponds to $\phi\to \tilde{\phi}$ 
in an infinite time $t_*$. Now $\phi$ has precisely enough kinetic energy to settle down in 
$\tilde{\phi}$ after an infinite time $t_*$. 

As for $Z=1$, substituting \eqref{Phit} in \eqref{phiPhib}, using \eqref{aa*b} and inverting 
\eqref{t*(t)b}, the Friedmann equations in GR for a spatially flat FLRW universe and for a 
minimally coupled scalar field with inverted potential \eqref{Vb} is completely integrated.

\section{Conclusions}
Viable bouncing solutions were found in the JF for a particular scalar-tensor gravity model 
equivalent to Einstein gravity with a cosmological constant and a conformally coupled scalar 
field with a quartic self interaction. Bouncing solutions were found for models with 
$\omega_{BD}>0$ and $-\frac{3}{2}<\omega_{BD}<0$. In this work we have studied in details in 
the EF all these bouncing solutions in the JF, whether these are viable or not. 
The various dynamical behaviours corresponding to bouncing universes in the JF are better
understood by inspection of the potential $V$ in the EF. Indeed, the two critical points 
for the late times dynamics of $\Phi$ have a direct physical meaning. The unstable viable 
bouncing solutions tending to $\tilde\Phi$ at $t\to \infty$ corresponds to solutions tending 
to the top of the potential $V$ at $\tilde\phi$. This is clearly an unstable fixed point which 
is reached for just one set of initial conditions. The other viable bouncing solutions tend to 
$\Phi=0$ and tend to the same value in the EF for $Z=1$. Inspection of $V$ shows clearly 
that a set of initial conditions for which $\phi$ passes the top of the potential with 
negative velocity will eventually settle down at $\phi=0$ after an infinite time $t_*$. 

Essentially the same analysis applies to the case $Z=-1$. 
The interval $0\le \phi < \infty$ in the EF covers the interval of physical values 
$0 \le \Phi < \Phi_{max} = \frac{\sqrt{6}}{\kappa}$ in the JF for $Z=1$. For $Z=-1$, the 
interval $0 < \phi < \infty$ covers the interval of physical values 
$\infty > \Phi > \frac{\sqrt{6}}{\kappa}$. 
When $\phi\to \infty$ in all cases a Big bang at negative times or a Big Crunch at positive 
times is obtained: this is how the unphysical limit $F\to 0$ manifests itself in the EF.   

For $Z=-1$, the unviable solution in the JF with $\Phi\to \frac{\sqrt{6}}{\kappa}$ in a finite 
time $t$ corresponds to a solution in the EF with $\phi\to \infty$, however this point is 
reached for $t_*=\infty$. Hence, the EF dynamics covers the physically acceptable part of 
the dynamics and it would be viable if the EF time $t_*$ would be the physical time. As we 
took the JF time $t$ as being the physical time, this solution was ruled out in \cite{BGPS15}. 
However if we view the EF dynamics for its own sake, independently of the underlying JF 
correspondence, this solution is perfectly regular in the future.   
 
We emphasize another important aspect of this work. From our study in the EF we get 
an integrable scalar field cosmological model where the singularities are standard 
cosmological singularities. Besides the interesting fact that a Big Crunch can be 
obtained for spatially-flat FLRW models, all these results are obtained for inverted 
potentials unbounded from below. We stress that solutions starting from a Big Bang and 
which are regular in the future and tending to a de Sitter space are obtained as well. 
Though we caution that these solutions should be viewed as interesting toy models and 
not as realistic cosmological universes, our results show nevertheless that inverted 
potentials unbounded from below can give interesting physical systems as it is 
the case here and in other bouncing \cite{OR13} or inflating \cite{IT12} models. 

%\section*{Acknowledgments}
%%%%%%%%%%%%%%%%%%%%%%%%%%%%%%%%%%%%%%%%%%%%%%%%%%%%%%%%%%%%%%%%%%%%%%%%%%%%%%%%%%%%%%%%%%%%%%
%%%%%%%%%%%%%%%%%%%%%%%%%%%%%%%%%%%%%%%%%%%%%%%%%%%%%%%%%%%%%%%%%%%%%%%%%%%%%%%%%%%%%%%%%%%%%%


\begin{thebibliography}{99}

\bibitem{Star78} A.~A.~Starobinsky, Sov. Astron. Lett. {\bf 4}, 82 (1978).

\bibitem{Page84} D.~N.~Page, Class. Quantum Grav. {\bf 1}, 417 (1984).

\bibitem{Ka98} A.~Yu.~Kamenshchik, I.~M.~Khalatnikov, A.~V.~Toporensky, 
 Int. J. Mod. Phys. D{\bf 7}, 673 (1997) [gr-qc/9801064].

\bibitem{GT08} G.~W.~Gibbons, N.~Turok, Phys. Rev. D {\bf 77}, 063516 (2008) 
 [hep-th/0609095].

\bibitem{DSNA11} D.~Tretyakova, A.~Shatskij, I.~Novikov and S.~Alexeyev, 
 Phys. Rev. D {\bf 85}, 124059 (2012) [arXiv:1112.3770].

\bibitem{BS13} J.~D.~Barrow, D.~Sloan, Phys. Rev. D {\bf 88}, 023518 (2013) 
 [arXiv:1304.6699].

\bibitem{APS06} A.~Ashtekar, T.~Pawlowski, P.~Singh, Phys. Rev. D {\bf 74}, 084003 (2006) 
 [gr-qc/0607039].

\bibitem{QECZ11} T.~Qiu, J.~Evslin, Y.~F.~Cai, M.~Li, X.~Zhang, 
 JCAP {\bf 1110}, 036 (2011) [arXiv:1108.0593].

\bibitem{ESV11} D.~A.~Easson, I.~Sawicki, A.~Vikman, JCAP {\bf 1111}, 021 (2011) 
 [arXiv:1109.1047]. 

\bibitem{BP14} D.~Battefeld, P.~Peter, Phys. Rept. {\bf 571}, 1 (2015) [arXiv:1406.2790]. 

\bibitem{BGPS15} B.~Boisseau, H.~Giacomini, D.~Polarski, A.~A.~Starobinsky, JCAP {\bf 0715}, 
 XXX (2015) to appear [arXiv:1504.07927].

\bibitem{BEPS00} B.~Boisseau, G.~Esposito-Far\`ese, D.~Polarski, A.~A.~Starobinsky, 
 Phys. Rev. Lett. {\bf 85}, 2236 (2000) [gr-qc/0001066].

\bibitem{GPRS06} R.~Gannouji, D.~Polarski, A.~Ranquet, A.~A.~Starobinsky, 
 JCAP {\bf 0609}, 016 (2006) [astro-ph/0606287].

\bibitem{MTZ03} C.~Martinez, R.~Troncoso, J.~Zanelli, Phys. Rev. D{\bf 67}, 024008 (2003) 
 [hep-th/0205319].

\bibitem{HPP06} S. de Haro, I. Papadimitriou, A. C. Petkou, Phys. Rev. Lett. {\bf 98}, 
 231601 (2007) [hep-th/0611315]. 

\bibitem{P13} D. Polarski, [arXiv:1303.4470].

\bibitem{AbrSte} M. Abramowitz and I. A. Stegun, Handbook of mathematical 
 functions with formulas, graphs, and mathematical tables (Milton Abramowitz and 
 Irene A. Stegun, eds.), Dover Publ., Inc., New York, 1992 (reprint of the 1972 edition).

\bibitem{Rub09} V.~A.~Rubakov, JCAP {\bf 0909}, 030 (2009) [arXiv:0906.3693].

\bibitem{BFG07} B.~Boisseau, P.~Forgacs, H.~Giacomini, J. Phys. A40, F215 (2007) 
[hep-th/0611306].

\bibitem{EP00} G.~Esposito-Far\`ese, D.~Polarski, Phys. Rev. D {\bf 63}, 063504 (2001)
 [gr-qc/0009034].

\bibitem{GV03} M.~Gasperini, G.~Veneziano, Phys. Rept. {\bf 373}, 1 (2003) 
 [hep-th/0207130].

\bibitem{FSS13} P.~Fr\'e, A.~Sagnotti, A.~S.~Sorin, Nucl. Phys. B{\bf 877}, 1028 (2013)
 [arXiv:1307.1910]. 

\bibitem{OR13} M.~Osipov, V.~Rubakov, JCAP {\bf 1311}, 031 (2013) [arXiv:1303.1221]. 

\bibitem{IT12} M.~M.~Ivanov, A.~V.~Toporensky, Int. J. Mod. Phys. D {\bf 21}, 
1250051 (2012) [arXiv:1112.4194].

\end{thebibliography}
\end{document}